\documentclass[useAMS,usenatbib]{mn2e}

\newcommand{\Msun}{\mbox{\,M$_\odot$}}

\newcommand{\mic}{\mbox{$\,\mu$m}}
\newcommand{\pion}[2]{{#1}\,{\sc {#2}}}

\newcommand{\ltsimeq}{\raisebox{-0.6ex}{$\,\stackrel
        {\raisebox{-.2ex}{$\textstyle <$}}{\sim}\,$}}
\newcommand{\gtsimeq}{\raisebox{-0.6ex}{$\,\stackrel
        {\raisebox{-.2ex}{$\textstyle >$}}{\sim}\,$}}

\newcommand{\XX}{\mbox{XX~Oph}}
\newcommand{\sirtf}{\mbox{\it Spitzer Space Telescope}}

\title[C$_{60}$ in \XX]{Solid-phase  C$_{60}$ in the peculiar binary \XX?}
\author[A. Evans et al.]{A. Evans$^{1}$\thanks{E-mail: ae@astro.keele.ac.uk},
J. Th. van Loon$^1$, C. E. Woodward$^2$, R. D. Gehrz$^2$, 
G. C. Clayton$^3$,\newauthor 
L. A. Helton$^{2,4}$, M. T. Rushton$^5$, S. P. S. Eyres$^5$, J. Krautter$^6$, S.
Starrfield$^7$,\newauthor  R. M. Wagner$^8$\\
$^{1}$Astrophysics Group, Keele University, Keele, Staffordshire, ST5 5BG, UK\\
$^2$Minnesota Institute for Astrophysics, School of Physics \& Astronomy,
116 Church Street SE, University of Minnesota, \\ Minneapolis, MN 55455, USA \\
$^3$Department of Physics and Astronomy, Louisiana State University, Baton
Rouge, LA 70803, USA \\
$^4$Stratospheric Observatory for Infrared Astronomy, NASA Ames Research Center,
MS 211-3, Moffett Field, CA 94035, USA\\
$^5$Jeremiah Horrocks Institute, University of Central Lancashire,
     Preston, PR1 2HE, UK \\
$^6$Landessternwarte, Zentrum f\"ur Astronomie der Universit\"at Heidelberg,
Koenigstuhl, D-69117 Heidelberg, Germany \\  
$^7$School of Earth and Space Exploration, Arizona State University, PO
Box~871404, Tempe, AZ 85287-1404, USA\\
$^8$Large Binocular Telescope Observatory, 933 North Cherry Avenue, Tucson, AZ
    85721, USA \\
}

\begin{document}

\date{Version of 2011-10-07}

\pagerange{\pageref{firstpage}--\pageref{lastpage}} \pubyear{2011}

\maketitle

\label{firstpage}

\begin{abstract}
We present infrared spectra of the binary \XX\ obtained with the Infrared
Spectrograph on the \sirtf. The data show some evidence for the presence
of solid C$_{60}$  -- the first detection of C$_{60}$ in the solid phase --
together with the well-known ``Unidentified Infrared'' emission features. We
suggest that, in the case of \XX, the C$_{60}$ is located close to the hot
component, and that in general it is preferentially excited by stars having
effective temperatures in the range $15\,000-30\,000$\,K. C$_{60}$ may be common
in circumstellar environments, but un-noticed in the absence of a suitable
exciting source.
\end{abstract}

\begin{keywords}
circumstellar matter --
astrochemistry --
stars: individual, \XX\ --
binaries: symbiotic --
infrared: stars
\end{keywords}

\section{Introduction}

The possible existence of buckminsterfullerene (C$_{60}$) in astrophysical
environments has long been suggested \citep{kroto}, but only recently has
observational evidence for emission from C$_{60}$ in the gas phase been
forthcoming. Gas phase C$_{60}$ has now been detected in  
the environments of young planetary nebulae \citep{cami,zhang}, RCB stars
\citep*{garcia} and in the reflection nebulae NGC\,2023 and NGC\,7023 that are
illuminated by B~stars \citep{sellgren}. In the case of the low-excitation
planetary nebula Tc~1, \cite{cami} argue that the C$_{60}$ (and
C$_{70}$) molecules are attached to the surfaces of cooler carbonaceous grains.
Many of the objects displaying C$_{60}$
also have strong ``Unidentified Infrared'' (UIR) features.

The formation of C$_{60}$ and other fullerenes in terrestrial laboratories
usually requires a hydrogen-deficient environment, and this seems to be
consistent with their presence in the environments of (evolved) H-deficient
carbon stars \citep{cami,garcia}. However the detection of C$_{60}$ in the
reflection nebulae NGC\,2023 and NGC\,7023 \citep{sellgren} indicates that
fullerene formation is possible in young (H-rich) environments. UIR features, as
well as ``Extended Red Emission'' attributed, among other hypotheses, to
small -- possibly ionised -- hydrocarbon molecules, are seen in the environment
of NGC\,7023 \citep{berne,sellgren}.

We report here the possible detection of solid phase C$_{60}$, in the
environment of the peculiar binary \XX, observed with the \sirtf\ 
\citep{wernera,gehrz}.

\section{The \XX\ binary}
 \label{binary}

\XX\ is a binary consisting of a late (M7III)
giant and an early (B0V?) star (see e.g. \citealt{dewinter, evansetal93} and
references therein; \citealt{cool} give M6-8II). It is sometimes classed as a Be
star \citep[e.g.][]{dewinter} and sometimes as a symbiotic \citep{GCVS}.
However it shows few of the common symptoms of symbiosis, such as the presence
of high excitation emission lines.

\begin{figure}
\setlength{\unitlength}{1cm}
\begin{center}
\leavevmode
\begin{picture}(5.0,6.)
\put(0.0,4.0){\includegraphics{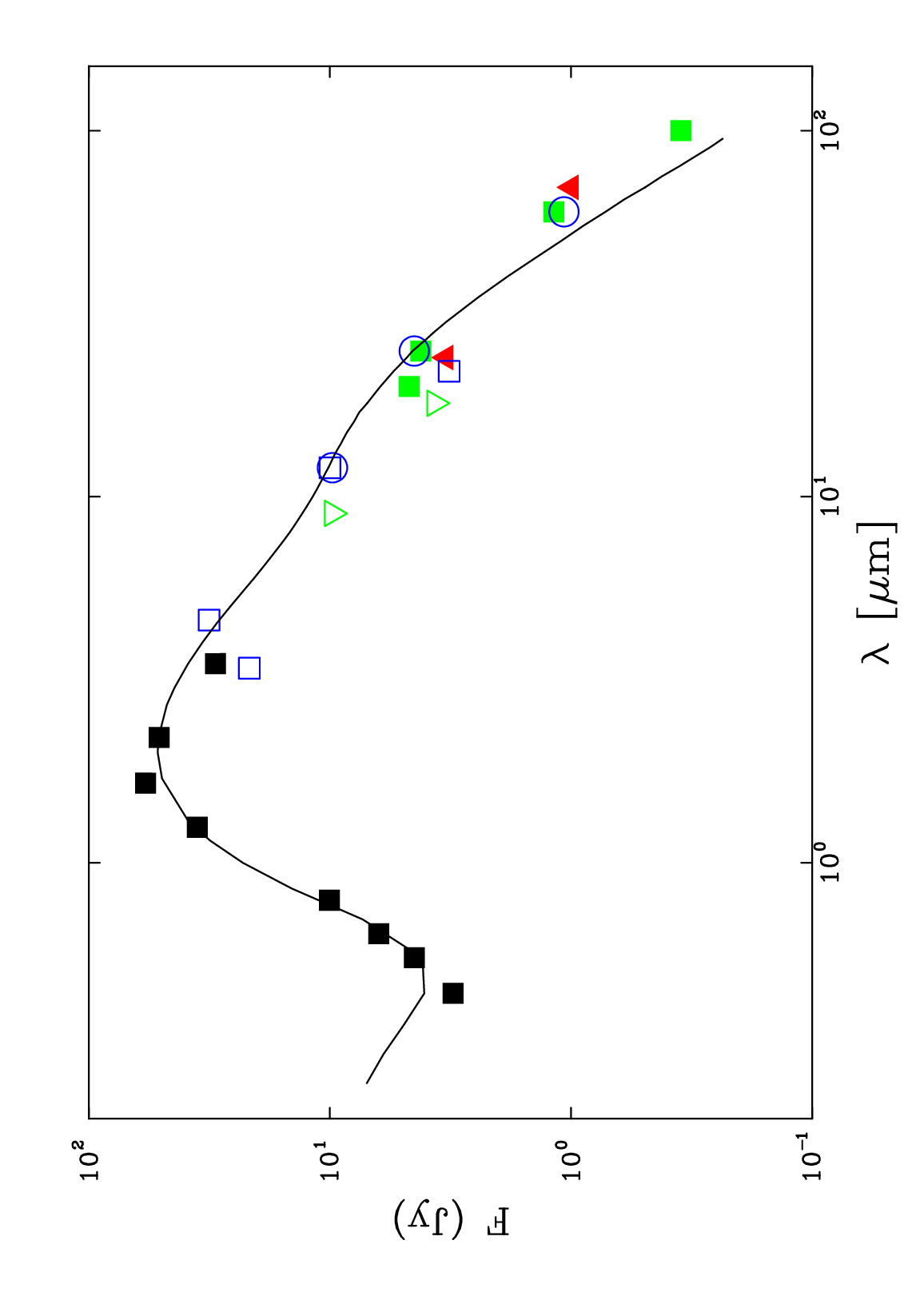}}
\end{picture}
\caption[]{Photometry of \XX, dereddened for $E(B-V)=0.51$ \citep{evansetal93}.
Filled black squares, $BV\!RI\,J\!H\!K\!L$ \citep{evansetal93};
open blue squares, WISE \citep{WISE};
inverted green triangles, AKARI \citep{AKARI};
open blue circles, IRAS PSC;
filled green squares, ISO PHOT-P;
red triangles, \sirtf\ MIPS.
In all cases errors are smaller than the plotted points.
Curve is DUSTY fit with parameters given in text; see text for details.
\label{SED}}    
\end{center}
\end{figure}

While there is photometric and spectroscopic evidence that a cool
component in the \XX\ system dominates in the red
\citep{dewinter,evansetal93,cool}, understanding the nature of the hot component has proven to be problematic. The evidence is circumstantial: there is
spectroscopic evidence for a `hot  companion' in the blue, in the form of H (and
other) emission lines.

\cite*{lockwood} argued that \XX\ is heavily reddened and estimated the
extinction, $A_{\rm v}$, to be $\sim4$~mag. Spectrophotometry of \XX\ was
presented by \cite{blair}, who deduced $E(B-V)\simeq 1.08$~mag on the basis of the
H$\alpha$/H$\beta$ ratio. They noted that this is significantly less than the
value given by \citeauthor{lockwood}, unless the ratio of total-to-selective
extinction is $R\simeq3.7$, but the polarization of \XX\ is inconsistent with a
high value of $R$ \citep{evansetal93}.

Although a B0V classification is assigned to the hot component, the presence of
a massive ($\sim20$\Msun) star in the \XX\ system seems unlikely on kinematic
grounds. For a distance of $\sim2$~kpc \citep{evansetal93} it lies $\sim400$~pc
above the Galactic plane and its proper motion \citep{hipparcos} takes it
towards the plane -- highly unlikely for a B0V star.

Furthermore, the spectral
energy distribution (SED) -- from 4400\AA\ to 100\mic\ -- can be fit (bearing in
mind the variability) by a two component DUSTY \citep{dusty} model. The hot component is a B {\em subdwarf} at the
centre of a dust shell having 0.01\mic\ amorphous carbon
grains with a temperature of 800~K at the inner boundary and an optical depth of
$\sim0.001$ in the visual. The cool component is a M7III star that effectively plays no part in heating the dust (see Fig.~\ref{SED}).
The DUSTY fit assumes that the dust shell is spherically symmetric with the
B~star located at its centre, so clearly the fit has its limitations (e.g. a
disc is more likely in a binary). However, the inner boundary of the dust shell
is $\sim7.2\times10^{11}$~m from the B~star. The size of the Str\"omgren
sphere associated with the B~star exceeds this if the gas density in its
vicinity $\ltsimeq10^{13}$~m$^{-3}$.

The reclassification of the hot component as a subdwarf removes the need for the
large reddening assigned by \citeauthor{lockwood} and others, and is consistent
with an interstellar reddening $E(B-V)=0.51$~mag \citep{evansetal93}. It also
has implications for the nature and evolution of the binary.

Although the cool component in \XX\ seems to be oxygen-rich -- as evidenced by
the presence of TiO and VO bands -- the 8.6\mic\ and 11.2\mic\ UIR features
reported in the IR spectrum of \XX\ by \citet{evans} are typical of carbon-rich 
environments. However the usual 3.28\mic\ and 3.4\mic\ UIR features are weak.

\begin{figure}
\setlength{\unitlength}{1cm}
\begin{center}
\leavevmode
\begin{picture}(5.0,5.)
\put(0.0,4.0){\includegraphics{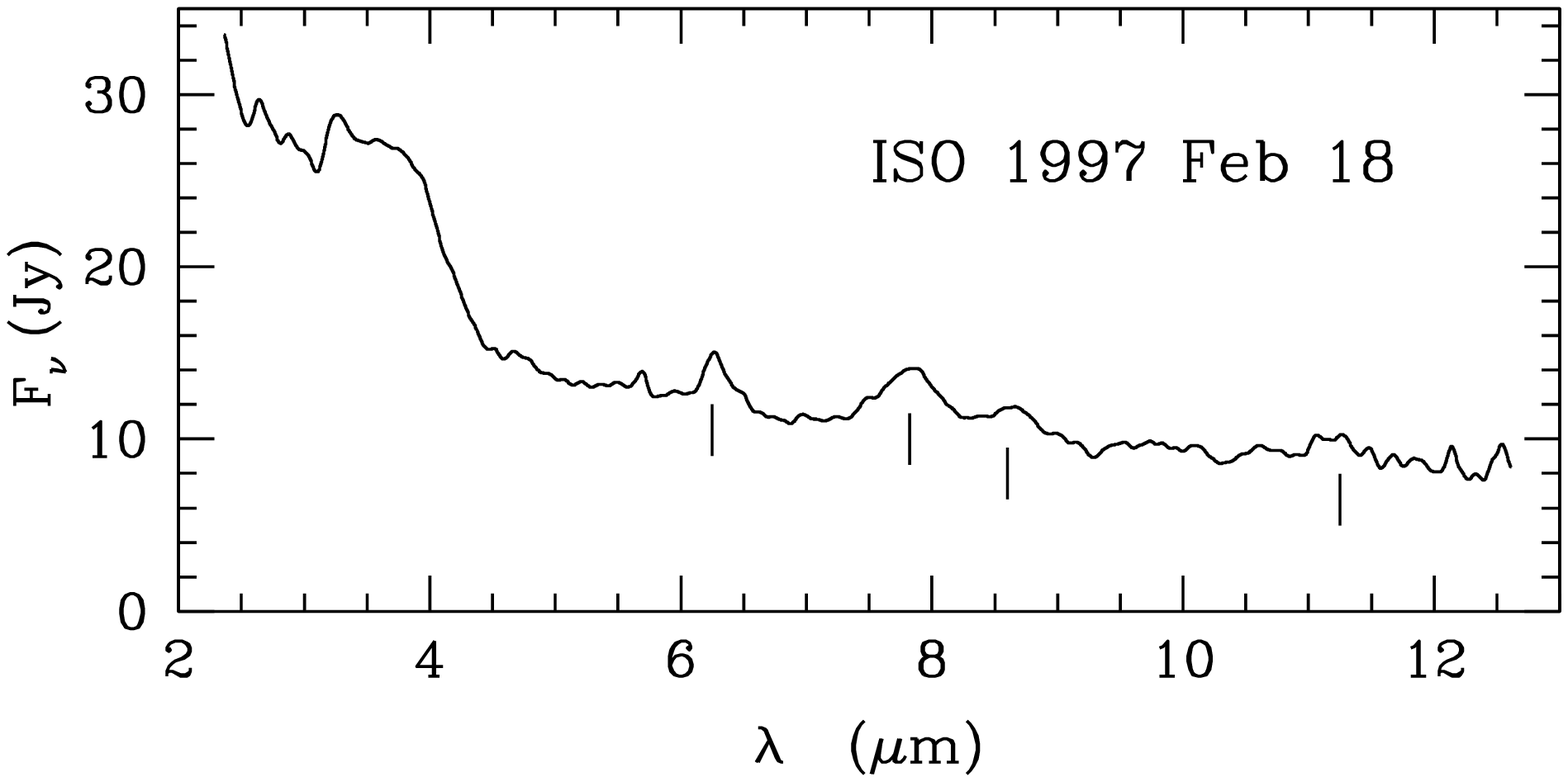}}
\end{picture}
\caption[]{ISO SWS spectrum of \XX; the 6.25\mic, 7.7\mic, 8.6\mic\ and
11.2\mic\ UIR features are identified. The apparent peak at 3.3\mic\ may be due
to the 3.3\mic\ UIR feature but the spectrum in this region is dominated by
molecular absorption in the M~giant.
\label{iso-sws}}
\end{center}
\end{figure}

Fig.~\ref{iso-sws} shows a spectrum of \XX\ obtained with the Short Wavelength
Spectrometer \citep[SWS;][]{degraauw} on the {\it Infrared Space Observatory}
\citep{kessler} that confirms the UIR features at 8.6\mic\ and 11.2\mic\
reported by \citet{evans}. The UIR feature at 6.25\mic\ is detected
and the non-detection of the 3.28\mic\ and 3.4\mic\ UIR features is confirmed.
The `8\mic' feature reported by \citeauthor{evans} is the long wavelength wing
of the well-known `7.7\mic' feature, affected by inadequate cancellation of the
atmosphere near the edge of the $8-13$\mic\ window. 

In most stars the flux in the 3.28\mic\ UIR feature is typically comparable to
that of the `7.7' feature \citep[e.g.][]{tielens}. On this basis we would expect
the 3.28\mic\ feature in \XX\ to have a peak flux $\sim3$\,Jy. While there is
indeed evidence for a feature at $\sim3.3$\mic\ (see Fig.~\ref{iso-sws}), the
spectrum in this region is dominated by molecular absorption (e.g. CO, OH) in
the M~giant, which has a flux of $\sim 32$\,Jy at 3\mic. The apparent
absence of the 3.28\mic\ feature can presumably be attributed to the fact that
it is swamped by the emission from the M~star.

\begin{figure}
\setlength{\unitlength}{1cm}
\begin{center}
\leavevmode
\begin{picture}(5.0,5.)
\put(0.0,4.0){\includegraphics{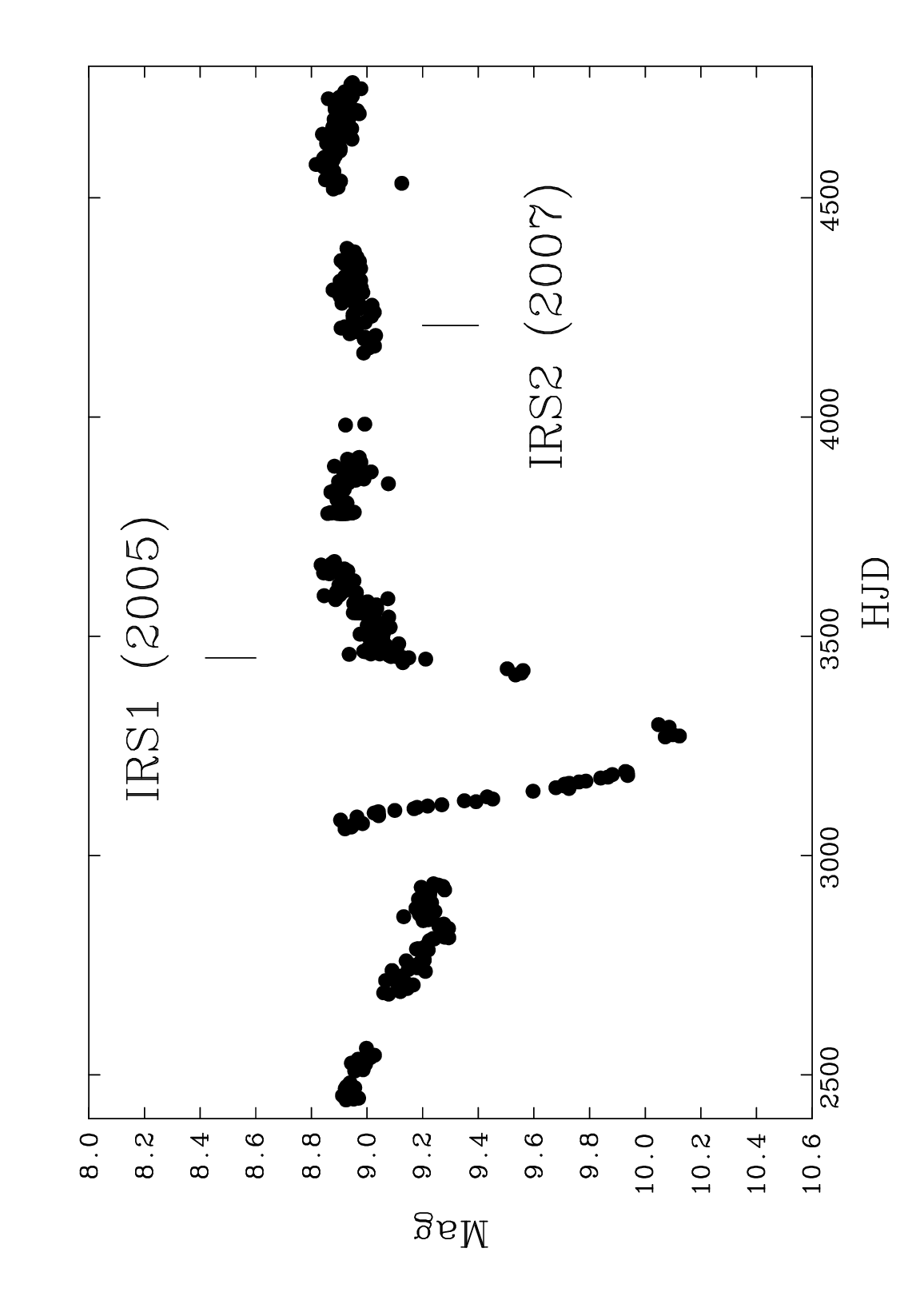}}
\end{picture}
\caption[]{The $V$ band lightcurve of \XX\ from the {\bf All Sky Automated
Survey} (ASAS) database \citep{ASAS}. The times of the \sirtf\ IRS observations
are indicated.}
\label{LC}
\end{center}
\end{figure}

The variability of \XX\ is irregular, although the Hipparcos catalogue
\citep{hipparcos} lists it
as having a possible period of 3.52~days, and as displaying sudden dips in 
luminosity. \cite{sobotka} reported that \XX\ went into a deep
(eclipse-like) minimum in 2005, the first for 37~years (see Fig.~\ref{LC}).
\cite{cool} found that the equivalent width of H$\alpha$ increased during the
minimum, indicating that the continuum around 656~nm had faded.

We note that, with a B subdwarf, most of the $V$-band light from the \XX\ system
comes from the M star, so that the eclipse in Fig.~\ref{LC} must be of the
giant, presumably by material in the vicinity of the B star. The optical depth
at $V$ at eclipse minimum is $\tau_V\simeq1.0$, far greater than that required
for the IR excess in Fig.~\ref{SED}, underlining the fact that the DUSTY
fit should not be taken too literally.

\section{Observations}

\XX\ was observed with the {\it Spitzer} Infrared Spectrograph
\citep[IRS;][]{houck} in staring mode on two occasions as \XX\ was emerging from
a deep minimum and some two years thereafter. The blue
peak-up array was used to centre the object in the IRS slits.
Observations were also obtained with the Multi-band Imaging Photometer for
{\it Spitzer} \citep[MIPS;][]{MIPS}. Spectra were obtained with both low- and
high-resolution IRS modes, covering the spectral range of 5--38\mic.
For the high-resolution modes we also obtained observations of the background;
however as we are comparing data from two epochs, the background measurement is
not critical. The spectrum was extracted from the version 12.3 processed
pipeline data product using SPICE version 2.2 \citep{spice}. 

\begin{figure*}
\setlength{\unitlength}{1cm}
\begin{center}
\leavevmode
\begin{picture}(5.0,9.)
\put(0.0,4.0){\includegraphics{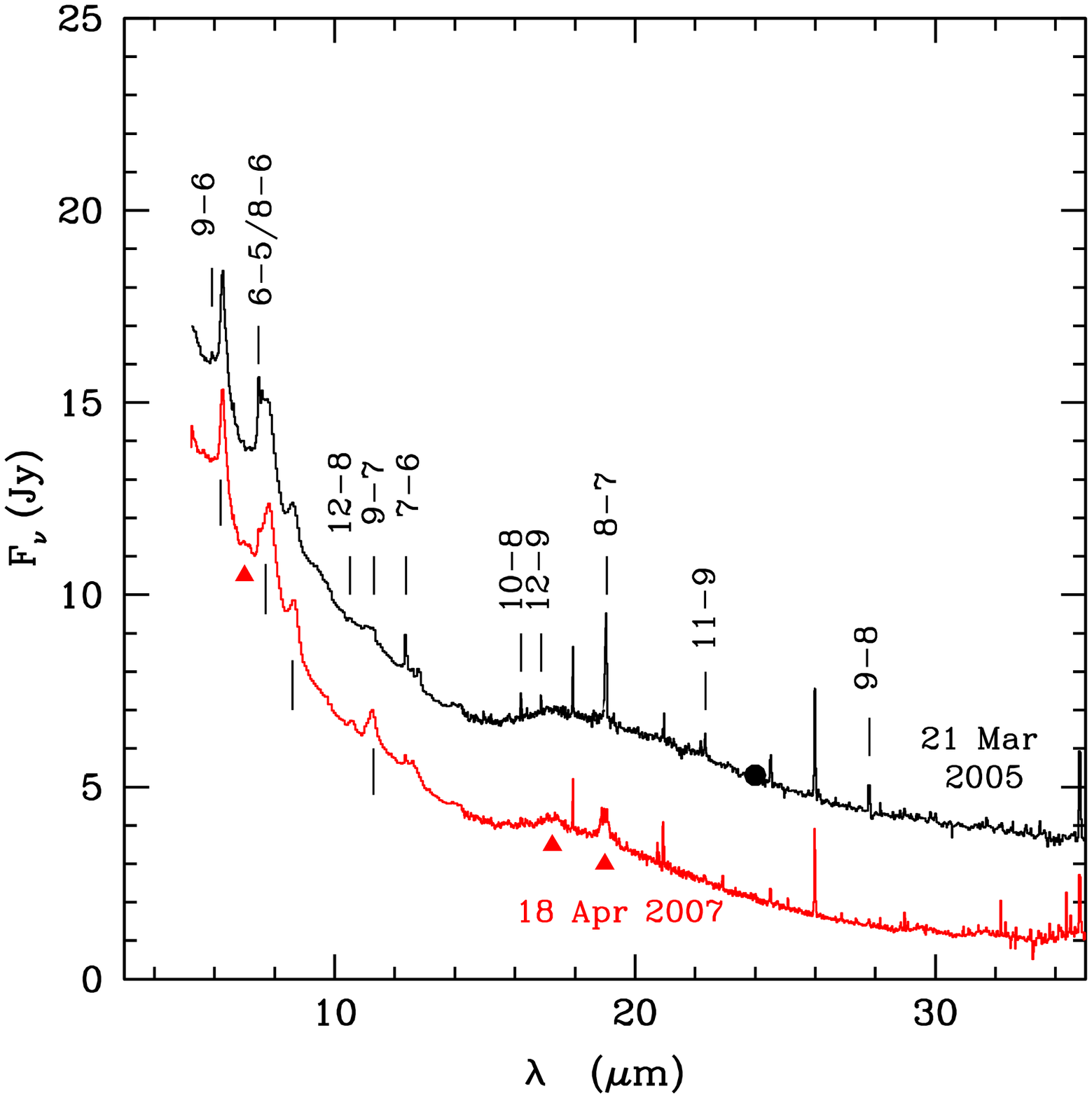}}
\end{picture}
\caption[]{\sirtf\ IRS spectrum of \XX\ in 2005 and 2007; the 2005 data have
been displaced upwards by 2~Jy for clarity. The 6.25\mic, 7.7\mic, 8.6\mic\ and
11.2\mic\ UIR and H recombination lines are identified; note the absence of H
recombination lines in 2007. The point at 24\mic\ is the MOPEX photometry. The
possible C$_{60}$ features at 7.1\mic, 17.25\mic\ and 19\mic\ in the 2007
data are indicated by triangles; see also Fig.~\ref{C60} below. \label{IRS1}} 
\end{center}
\end{figure*}

The spectra for the two epochs are shown in Fig.~\ref{IRS1}. There may be some
evidence for the 18\mic\ silicate feature, but the corresponding 9.7\mic\
feature is very weak. However the UIR features are clearly 
present, as is an excess longward of $\sim15$\mic\ due  to emission by
circumstellar dust (cf. Fig.~\ref{SED}). Such ``chemical dichotomy'' (i.e.
environments with a mix of C-rich and O-rich dust) is of course not uncommon
\citep[e.g.][and references therein]{clayton}.

There has clearly been a change in the infrared (IR) spectrum between 2005 and
2007. In particular H recombination lines are present in 2005 (as \XX\ was
emerging from eclipse) but were apparently weak in 2007; for example, the flux
in Hu\,$\alpha$\,12.371\mic\ was $1.61[\pm0.05]\times10^{-15}$~W~m$^{-2}$ in
2005, compared with $5.5[\pm1]\times10^{-16}$~W~m$^{-2}$ in 2007.
However, both H\,$\alpha$ and H\,$\beta$ are present in an optical spectrum of
\XX\ obtained on 2007 May 8 (within days of the 2007 IRS spectrum) by one of us
(LAH), as are a number of ``Diffuse Interstellar Bands'' (DIBs), which most
likely are of interstellar origin. These data will be presented in a future
paper (Helton et al., in preparation). 

\begin{figure}
\setlength{\unitlength}{1cm} 
\begin{center}
\leavevmode
\begin{picture}(5.0,6.)
\put(0.0,4.0){\includegraphics{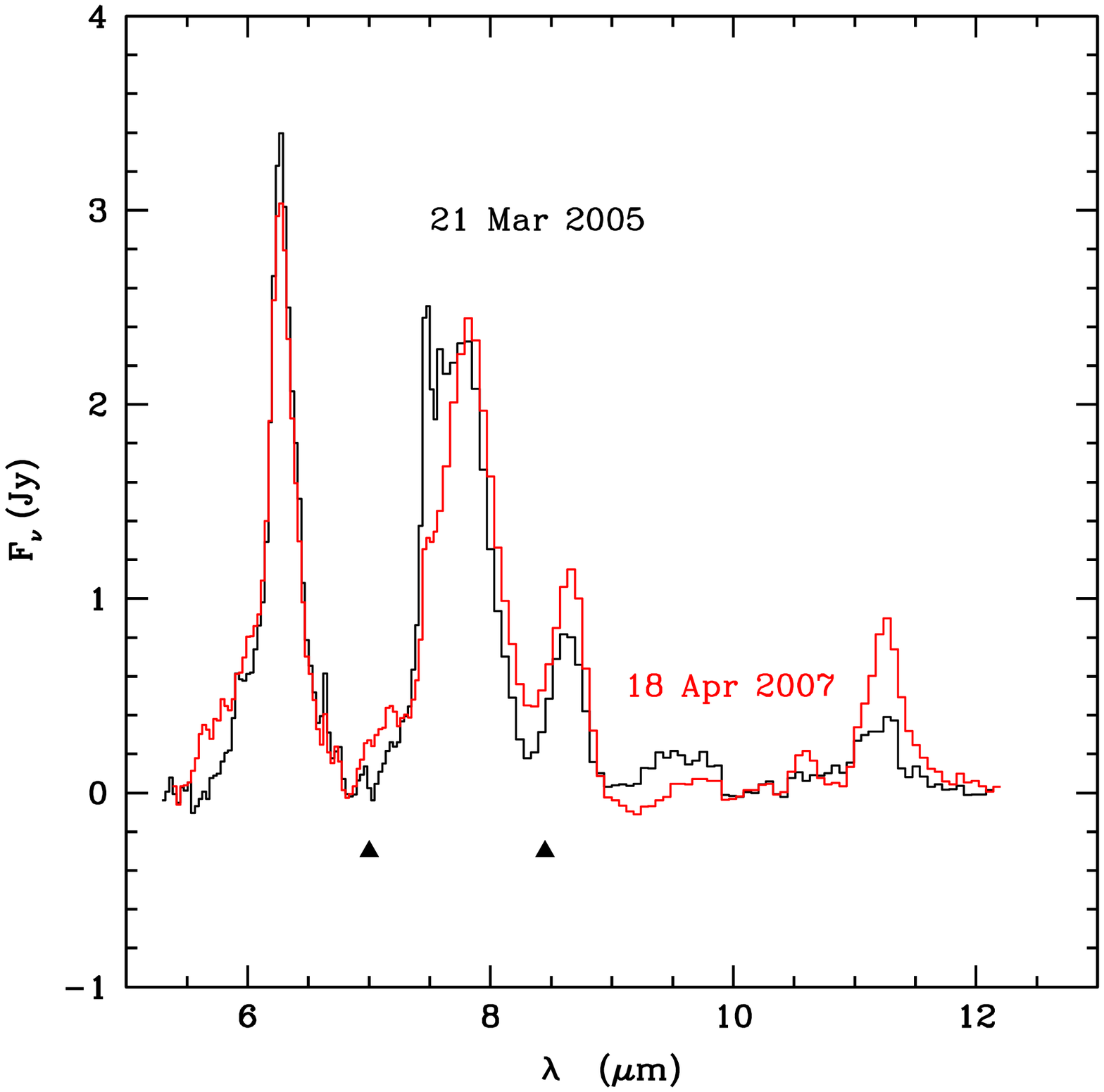}}
\end{picture}
\caption[]{UIR features in \XX; the expected wavelengths of the C$_{60}$
features are indicated by the triangles. The apparent ``excess'' at
$\sim7.5$\mic\ in the `7.7' UIR feature in 2005 is due to the presence of
\pion{H}{i} 6--5 and 8--6. The flux uncertainties in this wavelength range are
typically $\pm0.03$~Jy. \label{UIR1}} 
\end{center}
\end{figure}

We have extracted a continuum from both spectra to highlight the UIR features;
the result is shown in Fig.~\ref{UIR1}. There was little change between 2005 and
2007, except that the 11.3\mic\ and 8.6\mic\ UIR features were significantly
stronger in 2007 (when the IR hydrogen emission lines were weak).

The central wavelengths of the UIR features in \XX\ (e.g.
$7.48\pm0.01$\mic\ in 2005, $7.79\pm0.01$\mic\ in 2007 for the `7.7' feature)
are consistent with excitation by a source with an effective temperature in
excess of $\sim10^4$~K \citep[e.g.][]{sloan,acke}, and therefore with
excitation by the B~star. The changes we see in the strengths and possibly
central wavelengths of the UIR features may be associated with changes in the
ionisation of the PAH \citep[e.g.][]{draine}, possibly as a result of changes in
the extinction in the dust shell around the B~star.

\begin{figure}
\setlength{\unitlength}{1cm}
\begin{center}
\leavevmode
\begin{picture}(5.0,6.)
\put(0.0,4.0){\includegraphics{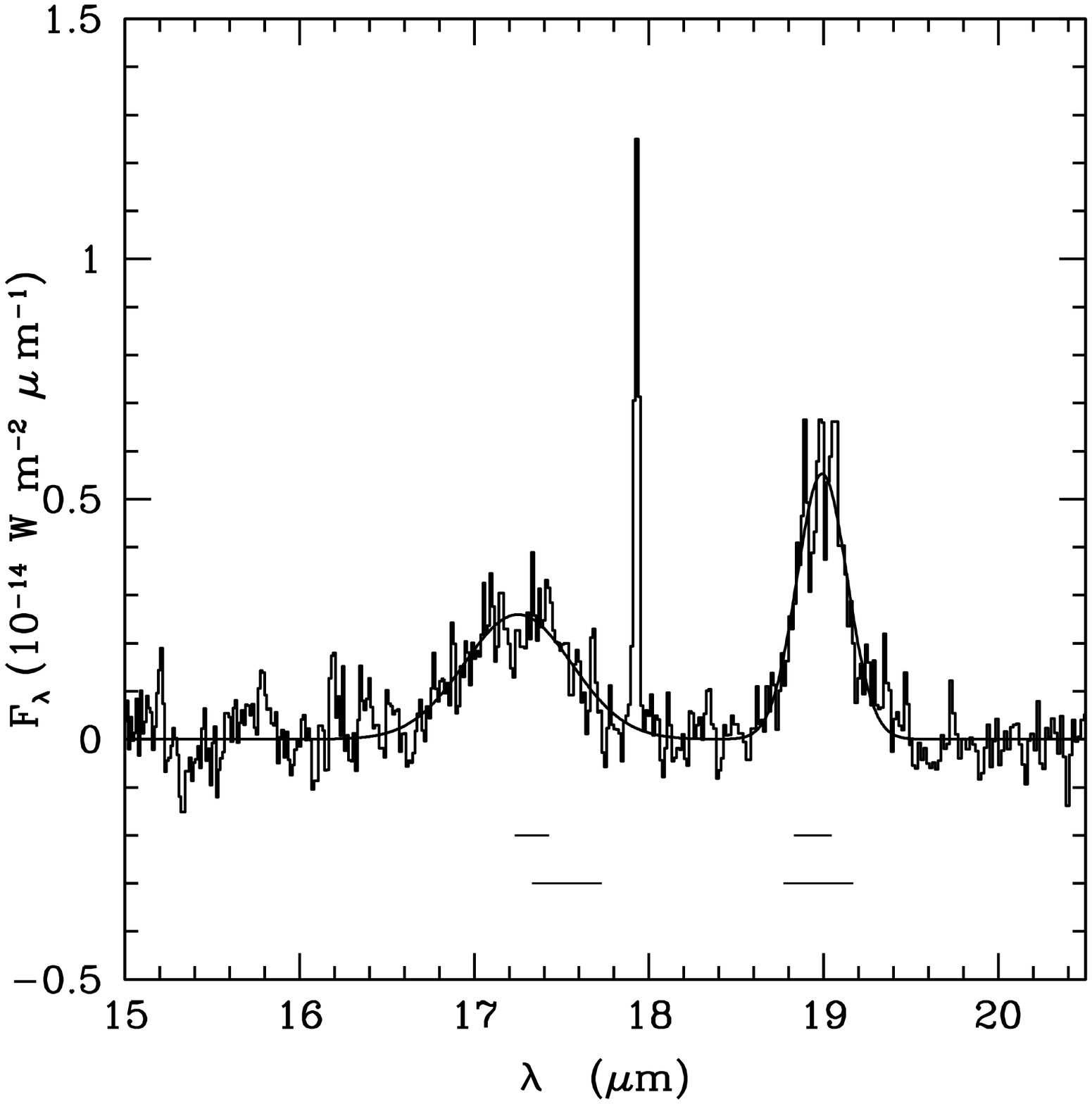}}
\end{picture}
\caption[]{Possible C$_{60}$ features in \XX. The upper horizontal lines
indicate the wavelengths of the ``17.3\mic'' and ``18.9\mic'' features in
C$_{60}$ smoke \citep{C60a}, the lower lines of the corresponding features in
gaseous C$_{60}$ \citep{C60b}. \label{C60}}
\end{center}
\end{figure}

Fig.~\ref{IRS1} also shows clear evidence for two broad features that are
present in 2007 but not in 2005. We have subtracted the continuum to highlight
these features (see Fig.~\ref{C60}). Features at these wavelengths are included
in the PAHFIT package \citep{pahfit} but we consider it unlikely that the two
features in \XX\ are due to emission by PAH molecules, for the following
reasons: (a)~their absence in 2005, when other UIR features were present;
(b)~the strength of the  19\mic\ feature compared with that of the 17.4\mic;
(c)~the 17.4\mic/7.7\mic\ flux ratio. The 17.4\mic\ feature was reported in
NGC\,7023 by \cite{wernerb}, who assigned it to ``aromatic hydrocarbons or
nanoparticles of unknown mineralogy''; it has subsequently been attributed to
C$_{60}$ \citep{sellgren}.

Possible identifications for these features are 17.4\mic\ and 18.9\mic\
C$_{60}$, which in the gas phase has four active vibrational
modes, at $\sim7.0\mic, 8.5\mic, 17.5\mic$ and 18.9\mic\ \citep[e.g.][]{C60b}.
Fig.~\ref{IRS1} also shows evidence for a feature at $\sim7$\mic, and
subtraction of the 2005 spectrum from the 2007 spectrum leaves a feature with
central wavelength $7.01\pm0.01$\mic\ (see Table~\ref{fluxes});
there is no evidence for the ``8.5\mic'' feature.

\begin{table*}
\begin{center}
\caption{Properties of C$_{60}$ features in \XX; wavelengths of gaseous and
solid C$_{60}$ features from Frum et al. (1991) and Kr\"atschmer et al. (1990)
respectively. Einstein coefficients $A$ from Mitzner \& Campbell (1995).}
\begin{tabular}{cccccc}  \hline
$\lambda$ ($\!$\mic) & FWHM ($\!$\mic) & Flux ($10^{-15}$~W~m$^{-2}$) & Gas $\lambda$
($\!$\mic) & Solid $\lambda$ ($\!$\mic) & $A$ (s$^{-1}$) \\ \hline
 $7.01\pm0.01$ &  $0.25\pm0.01$ &  $4.1\pm0.2$ & 7.11  & 7.00  & 151.6\\
 8.5   & &   $<0.2$    & 8.55 & 8.45 & 74.8 \\ 
$17.25\pm0.02$ & $0.72\pm0.04$ &  $2.00\pm0.10$ & 17.53 & 17.33 & 14.6 \\
$18.99\pm0.01$ & $0.34\pm0.01$ &  $2.00\pm0.10$ & 18.97  & 18.94 & 36.8 \\ \hline\hline
\end{tabular}
\label{fluxes}
\end{center}
\end{table*}

The ``17.4\mic'' feature we observe in \XX\ is actually at 17.25\mic,
quite different from the expected value of 17.53\mic\ for gas phase C$_{60}$
\citep{C60b}. However, {\it solid} C$_{60}$ has a feature at 17.3\mic\
\citep*{C60a,C60c}, closer to the 17.25\mic\ feature in \XX. We should
therefore consider whether the features in Fig.~\ref{C60} arise in gaseous or
solid C$_{60}$. 

The flux ratios of the putative C$_{60}$ features enable an estimate of the
vibrational temperature $T_{\rm vib}$ if the C$_{60}$ is in gaseous form. Using
Einstein coefficients from \citeauthor{mitzner} (1995; included in
Table~\ref{fluxes}), values of $T_{\rm vib}\sim520\pm50$~K are obtained; however
the ``18.9\mic'' flux seems underestimated by a factor $\sim2$. A similar value
($\sim670$~K) is obtained assuming that the energy of a $\sim10$~eV photon
absorbed by a C$_{60}$ molecule is equally distributed amongst the available
vibrational modes. However, at this temperature the ``8.5\mic'' feature would
have a flux $\sim1.5\times10^{-15}$~W~m$^{-2}$, far greater than observed. We
also note that laboratory measurements on solid C$_{60}$ \citep{C60a} suggest
that the ``8.5\mic'' feature is rather weaker than the other three. Therefore on
the basis of (i)~the wavelength of the ``17.4\mic'' feature and (ii)~the
weakness of the ``8.5\mic'' feature, we conclude that the C$_{60}$ in \XX\ is
most likely in solid form; if so this is the first astrophysical detection of
solid C$_{60}$.

The absorption cross-section of C$_{60}$ has been measured by \cite{yagi}, from
which we estimate the Planck mean absorption cross-section per C$_{60}$ molecule
(averaged over the emission of the B~star) to be $\sim7\times10^{-21}$~m$^2$.
If (cf. Section~\ref{binary}) the B~star is situated at
$\sim7.2\times10^{11}$~m from the inner boundary of the dust shell, 
the temperature of a C$_{60}$ grain of radius $a$ is
$\sim200\,(a/0.03\mic)^{1/4}$~K.

While the apparent absence of C$_{60}$ in the IRS spectrum immediately after
eclipse in 2005, and its presence in 2007, is suggestive, it is difficult to
argue that the eclipse is in any way connected with the presence of C$_{60}$ in
the spectrum, especially as it is the giant that is eclipsed: it is likely
therefore that the appearance of C$_{60}$ in 2007 is unconnected with the
eclipse of Fig.~\ref{LC}.

\section{C$_{60}$ in \XX}

We can make an estimate of the mass of C$_{60}$ using the combined flux in the
C$_{60}$ features. Assuming (cf. Section~\ref{binary}) the B~star is 
situated at $\sim7.2\times10^{11}$~m from the inner boundary of the dust shell,
and using the Planck mean absorption cross-section above,
the absorbed power per C$_{60}$ particle is $\sim8.1\times10^{-18}$~W. The
emitted power \citep[assuming a distance of 2~kpc for \XX;][]{evansetal93} is
$\sim3.9\times10^{26}$~W, so $\sim4.8\times10^{43}$ C$_{60}$ particles
(i.e. $\sim2.9\times10^{-11}$\Msun), in solid form, are required.
This suggests that the number of C$_{60}$ molecules is $\sim0.03$ the 
number of PAH molecules.

The detection of C$_{60}$ in a range of environments
\citep{cami,garcia,sellgren}, including both H-poor and H-rich environments,
indicates that C$_{60}$ can form in a variety of astrophysical conditions.
\citeauthor{garcia} suggest that both the UIR carrier and C$_{60}$ may form as a
result of the disintegration of hydrogenous amorphous carbon (HAC) grains.
However the fact that HAC is seen in environments \citep[e.g.
novae; cf.][]{evans2} in which C$_{60}$ is {\em not} seen indicates that there are
other factors that determine whether or not C$_{60}$ is detected.

Most of the objects in which C$_{60}$ has been reported are associated with
stars having effective temperature $T_{\rm eff}$ in the range
$\sim15\,000-30\,000$\,K, the exception being the RCB star V854~Cen ($T_{\rm
eff}\simeq6\,750$\,K). \XX\ is in the former category, while classical novae
have T$_{\rm eff}\gtsimeq50\,000$\,K at the time of dust formation. 
Notwithstanding the small number of objects in which C$_{60}$ has been detected,
the data thus far may point to the fact that it is the effective temperature of 
the central star that is the common factor in the detection of C$_{60}$, the
critical range being $\sim10\,000-30\,000$\,K.

In 2007 the C$_{60}$ in \XX\ seems to be present when the IR H recombination
lines are weak, and the 8.5\mic\ and 11.2\mic\ UIR features are strong. This
suggests either (a)~ that the C$_{60}$ is not a permanent feature of the \XX\
environment but is formed when conditions are favourable (either by
fragmentation of larger particles, or by chemical routes from smaller
molecules), or (b)~that C$_{60}$ is a permanent feature and that its excitation
is intermittent.

One possible scenario is that the C$_{60}$-bearing material is, as already
discussed, confined to the vicinity of the B star. The H lines arise from a
shell, also associated with the B star and possibly accreted from the giant wind;
the relative sizes of the ionised and dusty regions
depend on the gas density. The formation of C-rich dust would require the
photodissociation of wind CO by UV radiation from the B star to release C for
carbon chemistry \citep{evans}. Enhanced formation of C$_{60}$ (coincidentally
after the 2005 eclipse) would be consistent with the appearance of C$_{60}$ in
2007. Quenching of the UV radiation from the B star by the C$_{60}$-containing
dust would lead to reduced excitation of H in the shell. If this is correct then
it is likely that C$_{60}$ is formed ``bottom up'' rather than ``top down''. 

\section{Conclusions}

We have reported the possible detection of solid-phase C$_{60}$ in the
environment of the peculiar binary \XX. Contrary to previous work we conclude
that the hot star is a B subdwarf that is surrounded by an ionised shell and a
C$_{60}$-bearing shell, most likely in the form of a disc. Variations in the
optical depth of the latter results in variations in the excitation of H lines.

We will present a detailed discussion of the \XX\ system and its environment in
a forthcoming paper.

\section*{Acknowledgments}

We thank Dr L. d'Hendecourt for helpful comments on an earlier version.

This work is based on observations made with the \sirtf, which is operated by
the Jet Propulsion Laboratory, California Institute of Technology under a
contract with NASA.
This publication makes use of data products from the Wide-field Infrared Survey
Explorer, which is a joint project of the University of California, Los Angeles,
and the Jet Propulsion Laboratory/California Institute of Technology, funded by
the National Aeronautics and Space Administration.
Based on observations with AKARI, a JAXA project with the participation of ESA.
RDG, CEW and LAH were supported by various NASA {\it Spitzer}/JPL contracts and
the United States Air Force.
SS was supported by NASA and the NSF.

\appendix

\bsp

\label{lastpage}

\end{document}